\documentclass[sn-mathphys-num]{sn-jnl}

\usepackage{url} 
\usepackage{graphicx}%
\usepackage{multirow}%
\usepackage{amsmath,amssymb,amsfonts}%
\usepackage{amsthm}%
\usepackage{mathrsfs}%
\usepackage[title]{appendix}%
\usepackage{xcolor}%
\usepackage{textcomp}%
\usepackage{manyfoot}%
\usepackage{booktabs}%
\usepackage{algorithm}%
\usepackage{algorithmicx}%
\usepackage{algpseudocode}%
\usepackage{listings}%
\usepackage{subcaption}
\usepackage{arydshln}

\usepackage{hyperref}
\usepackage{caption}

\def\beq{\begin{equation}}
\def\eeq{\end{equation}}
\def\bea{\begin{eqnarray}}
\def\eea{\end{eqnarray}}
\def\figref#1{Fig.~\ref{fig:#1}}
\def\figlab#1{\label{fig:#1}}  
\def\tabref#1{Table~\ref{tab:#1}}
\def\tablab#1{\label{tab:#1}}  
\def\eqref#1{Eq.~(\ref{eq:#1})}

\def\eqlab#1{\label{eq:#1}}
\newcommand*{\secref}[1]{Section~\ref{sec:#1}}
\newcommand*{\seclab}[1]{\label{sec:#1}}
\newcommand*{\appref}[1]{Appendix~\ref{app:#1}}
\newcommand*{\applab}[1]{\label{app:#1}}
\renewcommand{\Im}{\operatorname{Im}}
\renewcommand{\Re}{\operatorname{Re}}

\def\ATRID{ATRI-D}

\begin{document}

\title{Measuring the locations and properties of VHF sources emitted from an aircraft flying through high clouds
}
\author*[1,2]{\fnm{Olaf} \sur{Scholten}}\email{o.scholten@rug.nl}

\author[1,2]{\fnm{Marten} \sur{Lourens}}
\author[3]{\fnm{Stijn} \sur{Buitink}}
\author[4]{\fnm{Steve} \sur{Cummer}}
\author[5]{\fnm{Joe} \sur{Dwyer}} %
\author[1,2]{\fnm{Brian M.} \sur{Hare}}
\author[6]{\fnm{Tim} \sur{Huege}} %
\author[5]{\fnm{Ningyu} \sur{Liu}}
\author[7]{\fnm{Katie} \sur{Mulrey}}
\author[8,9]{\fnm{Anna} \sur{Nelles}}
\author[2]{\fnm{Chris} \sur{Sterpka}}
\author[10]{\fnm{T.~N.~Gia} \sur{Trinh}}
\author[1,2]{\fnm{Paulina} \sur{Turekova}}
\author[2]{\fnm{Sander} \sur{ter Veen}}

\affil[1]{Kapteyn Astronomical Institute, University of Groningen, P.O. Box 72, 9700 AB Groningen, Netherlands}
\affil[2]{Netherlands Institute for Radio Astronomy (ASTRON), Dwingeloo, The Netherlands}
\affil[3]{Interuniversity Institute for High-Energy, Vrije Universiteit Brussel, Pleinlaan 2, 1050 Brussels, Belgium}
\affil[4]{Department of Electrical and Computer Engineering, Duke University, Durham, NC, 27708, USA}
\affil[5]{Department of Physics \& Astronomy and Space Science Center (EOS), University of New Hampshire, Durham NH 03824 USA}
\affil[6]{KIT, Karlsruhe, Germany}
\affil[7]{Department of Astrophysics/IMAPP, Radboud University Nijmegen, Nijmegen, The Netherlands}
\affil[8]{Deutsches Elektronen-Synchrotron DESY, Platanenallee 6, 15738 Zeuthen, Germany}
\affil[9]{Erlangen Centre for Astroparticle Physics (ECAP), Friedrich-Alexander-Universit\"{a}t Erlangen-N\"{u}rnberg, 91058 Erlangen, Germany}
\affil[10]{Physics Education Department, School of Education, Can Tho University, Campus II, 3/2 Street, Ninh Kieu District, Can Tho City, Viet Nam}


\abstract{
We show that it is possible to locate the few places on the body of an airplane, while it is flying through high clouds, from which broad-band, pulsed, radiation is emitted at Very High Frequency (VHF) radio frequencies. This serendipitous discovery was made whilst imaging a lightning flash using the Low-Frequency Array (LOFAR). This  observation provides insights into the way the airplane sheds the electrical charge it acquires when flying through clouds.
Furthermore, this observation allowed us to test and improve the precision and accuracy for our lightning observation techniques.

Our new results indicate that with the improved procedure the location precision for strong pulses
is better than 50~cm, with the orientation of linear polarization being accurate to within 25$^\circ$. For the present case of a Boeing 777-300ER, VHF emissions were observed exclusively associated with the two engines, as well as a specific spot on the tail. Despite the aircraft flying through clouds at an altitude of 8~km, we did not detect any emissions from electrostatic wicks.
}


\maketitle

\section{Introduction}

While examining a radio image of a lightning flash recorded by the LOFAR radio telescope on 2019-04-24 at 19:44:32~UTC, we observed a series of sources that appeared to trace an object moving slowly, at approximately 800~km/h (or 220~m/s), compared to the much higher propagation speeds typically associated with lightning discharges, which range from around $10^5$ to over $10^7$~m/s. Given that these sources were located at an altitude of about 8~km, it became clear that they originated from an airplane, as was confirmed by flight data. The sources exhibited neither regular timing nor consistent intensity, ruling out the possibility that they were emitted by a beacon. As the aircraft was flying through clouds, composed of frozen water at this altitude, where friction leads to charging of the aircraft~\cite{Martell:2022}, we expect that the VHF radio emissions were likely a result of electric discharge, although the positioning of the strongest sources on the airplane seem to disfavor this explanation. Electrostatic discharges tend to occur from the sharpest points on the airplane~\cite{Guerra-Garcia:2018} (e.g., the wings~\cite{Spark:2020}) which is not what we see.

Airplanes have been detected during lightning observations, as reported in~\cite{Zhang:2004}.
Also large astrophysical radio observatories have observed broad-band, short duration, radio pulses from airplanes, where a first reporting is given in~\protect{\cite{Wucknitz:2011}}. The observation of such relatively small sources are often very valuable for calibrating a large system of antennas such as used in astrophysical observations~\protect{\cite{Aab:2016, RNO-G:2025-plane, Agarwal:2025}}. On the other hand, for some other astrophysical observations they are a potential nuisance~\protect{\cite{Gehlot:2024, Ducharme:2025}}. Other sources of detectable broad-band signals include reflections of DAB/DTV signals from aircraft and meteorite trails in the atmosphere~\protect{\cite{Tingay:2020}}, or from space objects in low Earth orbits~\protect{\cite{Prabu:2022}}. Again, this may be regarded as a background, or rather as an opportunity to use the DAB signal as a passive radar for detecting aircraft~\protect{\cite{Klos:2021, Jedrzejewski:2024}}.
None of these previous observations were able to determine the location of the emission region or regions within the airplane. We report here, for the first time, observations which can localize the emission to well within 1~m in the moving frame of the aircraft.
Using our near-field beamformer (where we coherently add the signals of about 200 dual-polarized antennas) we can even detect, and localize, weak broad-band, impulsive, sources on the plane that, for a single antenna, are barely visible above the background, which is mostly from galactic origin.
Additionally, we determine the 3-d polarization direction of the sources that emit the radiation.

This serendipitous finding opened up a number of avenues for further exploration:
\begin{enumerate}
\item For lightning imaging the detection of emissions from tightly confined sources is particularly valuable, as this allows for investigating the intrinsic accuracy of the imaging process.
\item The emission mechanism of the VHF emitting process in natural lightning is complex because of the inherently chaotic nature of lightning events. Theoretically the emission process remains largely speculative~\cite{Moss:2006, Kochkin:2012, Luque:2014, Kochkin:2015, Shi:2016, Shi:2019, Hare:2020, Liu:2021}. The observation of an electrostatic discharge from a more confined source may help to resolve these speculations.
\item The phenomenon of static electric charge emissions from airplanes appears to be relatively unexplored, and observations of VHF-emission due to electrical discharges may help enhance understanding.
\end{enumerate}

LOFAR~\cite{Haarlem:2013} is a radio telescope consisting of thousands of antennas spread across much of Europe. In the observations we discuss here, we focus on the antennas located at the Dutch stations, specifically those operating in the 30 -- 80 MHz VHF band. Of particular importance for the observations presented in this work is the fact that the timing for each station is determined by an atomic clock which results in a timing stability of better than $10^{-9}$ (less than 1 nanosecond walk per second), that the stations are connected by a dedicated glass fiber network allowing for the storage of the 0.3 terabyte data volume per few seconds of observation, and that the antenna stations are spread over an area of a several thousand km$^2$. We use the antenna buffers (transient buffer boards) to save raw-voltages and perform all processing offline since, for the short pulses we are interested in, we cannot use the station beamformers or correlators used in astronomy observations.

The imaging methods we employed are based on the techniques described in Ref.~\cite{Scholten:2021-init}, utilizing a time-of-arrival-difference approach (the impulsive imager), as well as the time resolved interferometric 3-dimensional (TRI-D) procedure using interferometric beamforming. Since near-field beamforming of a polarized broad-band signal is not standard, we have summarized the essentials of our approach in \protect{\appref{TRID}}, while a more detailed discussion can be found in Ref.~\cite{Scholten:2021-INL}.  The combination of LOFAR with these imaging techniques enables us to capture the intricate details of lightning discharges, leading to the discovery of several new structures, such as needles~\cite{Hare:2019, Hare:2021}, intensely radiating negative leaders~\cite{Scholten:2021-RNL}, and high altitude negative leaders~\cite{Scholten:2021-HANL}. Simulations suggest that our spatial resolution is good, on the order of 10~cm~\cite{Scholten:2022}, but it would be beneficial to confirm this in actual observational data. The observation of charge emissions from an airplane thus presents a unique opportunity, as the point of emission is relatively localized, whereas the extent of the region from which VHF emissions occur in actual lightning events remains largely unknown on scales below 10~meters, and sometimes even up to 100~meters.

\begin{figure}[h]
\centering	
   \includegraphics[width=0.47\textwidth]{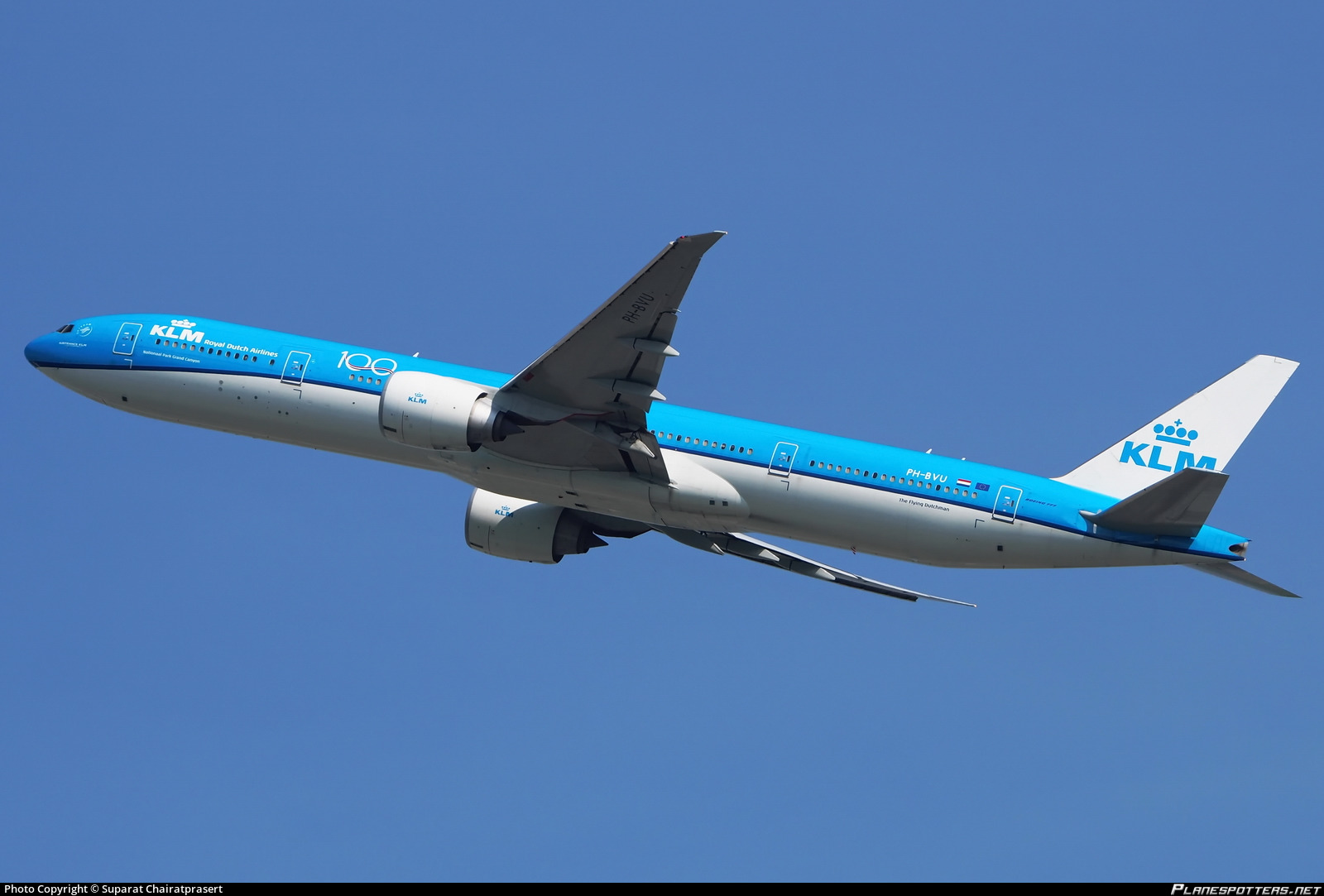}
	\caption{The Boeing 777-306 (ER) named "Grand Canyon National Park" we have observed with LOFAR (photo by Suparat Chairatprasert).}
	\figlab{PH-BVU}
\end{figure}

The aircraft we observed, a Boeing 777-306 (ER), turned out to be the, at that time, newest B77W plane on the KLM fleet~\cite{KLM-Fleet} and is shown in \figref{PH-BVU}. When planes fly through high clouds, electrostatic charge accumulates~\cite{Martell:2022}, which can pose serious safety risks, as it may trigger a lightning discharge from the aircraft~\cite{Guerra-Garcia:2018, Spark:2020}. As this involves large, quickly changing currents on the nano-second scale, one would expect that copious amounts of VHF radiation are emitted in this process.
Surprisingly, the observation presented in this work only shows VHF emission from sources associated with the engines, and the tail of the airplane, even while it was traversing through clouds.
We had anticipated observing discharges from the static dischargers, or p-static wicks, but none were recorded.
It is established that engine exhaust plays a crucial role in dissipating the aircraft's charge~\cite{Jones:1990, Martell:2022}, which might explain some of the VHF emissions we observed.

We have searched other LOFAR-lightning recordings for which we have made high-quality images (order twenty) for similar observations but found only one other recording in which we detected an airplane event. A first analysis of this second case is presented in the supplementary information. Although one expects planes to be present for every recording in the area that is imaged accurately we have only two recordings where we found an airplane. The reason for this is two-fold. Few planes are emitting strong signals that can be detected by our impulsive imager. Another reason is that the signals are weaker than the stronger lightning pulses and, have a larger spread in time hampering accurate imaging. The plane we observed in a recording of August, 14, 2020, flying though high clouds at an altitude of 11.6~km shows very similar features as the one detailed in this work (i.e., brief VHF radio frequency pulses form one location in the tail and the front and back of the two engines), whereas another plane in the same recording, flying at an altitude of about 12~km, well above the local cloud coverage, completely escaped our detection.

The observation that VHF emissions from the aircraft occur at only a few, very localized spots, led us to enhance our imaging pipeline. We further developed our procedure based on the beamforming technique utilized in the TRI-D imager~\cite{Scholten:2021-INL}. We discovered that a significant source of inaccuracies in the localization of these emissions stemmed from the arbitrary slicing of the time trace. Therefore, a clear improvement to our methodology is found to be the implementation of a dynamic windowing technique in the beamforming algorithm that captures the entire duration of the VHF pulse.

In \secref{Methods}, we provide a general overview of the methodology behind LOFAR lightning observations before narrowing our focus to the specific observation pertinent to this study. In \secref{ATRID}, we outline the method we employed to enhance the accuracy of our beamforming imager, TRI-D. The findings from the precise airplane observations are detailed in \secref{Results}, where we demonstrate that the absolute location accuracy of our observations is approximately 10~m, with a relative accuracy of about 50~cm for the emissions detected near the plane's tail.
Additionally, the observed sources exhibit a distinct polarization direction, as discussed in \secref{Polarization}. We determine a spread in the linear polarization direction of 25$^\circ$ which is a convolution of the intrinsic accuracy of the system and the physical spread in polarization of the emitting sources. 
Our observations also provide insight into the temporal structure of each VHF source, as explored in \secref{TimeTrace}, revealing that many begin with a strong, brief impulse followed by several weaker signals.
Some speculations are presented in \secref{Physics} on the interpretation of the observed sources and their distributions.
Finally, we summarize our conclusions in \secref{Conclusions}.

\section{Methods}\seclab{Methods}

In this work, the primary observation method is the same as presented in \protect{Ref.~\cite{Hare:2019}}. A source at a location $\vec{r}_s$ emits a short impulsive signal at time $t_s$ that is detected by a large number of antennas at locations $\vec{r}_a$. Ignoring, for ease of notation, the index of refraction, the arrival times of the signal in each antenna is $t_a=t_s +  |\vec{r}_a-\vec{r}_s|/c$ where $c$ is the light velocity. From a measurement of the pulse arrival time in antennas that are spread over a sufficiently large area thus the source location can be deduced.
Important for an accurate determination of the source location is thus that the antennas are spread over a large area (six dual polarized antennas from each of the 36 Dutch LOFAR stations, totalling to about 4000~km$^2$ for the present observations).
Arrival time difference can be determined to an accuracy of about 1 nanosecond for antennas that are at a distance of 100~km.
Essential to achieve this is the time stability of the LOFAR antennas combined with an accurate time calibration, see~\protect{\cite{Scholten:2021-init}} as well as true broad-band observations where we use the full 30 to 80~MHz band offered by LOFAR where the full resolution time traces are stored on the transient memory coupled to each antenna, for later off-line processing.

Lightning emits an enormously large number of short pulses (an almost flat frequency response in our bandwidth)  while propagating in the atmosphere. On one hand this allows for the accurate imaging of lightning development but it also poses a challenge to the imaging procedure. To address this issue we have developed two different imaging procedures. One we name the `impulsive' imager, discussed in detail in  \protect{Ref.~\cite{Hare:2019}}, which relies on observing arrival time differences of individual pulses where the pulse-confusion problem is minimized by using a Kalman-filter inspired approach~\protect{\cite{Scholten:2021-init}}. The other, named `TRI-D', voxelates the part of the atmosphere where sources are searched and employs a beamforming technique. A more detailed description of TRI-D is given in \appref{TRID}, see also~\protect{\cite{Scholten:2022}}. While the impulsive imager comprehensively maps the whole area where LOFAR can observe lightning discharges (an area of about $140\times 140$~km$^2$) the TRI-D imager can locate more detailed structures but operates in tesseracts of typically 150~m~$\times$ 150~m~$\times$ 150~m~$\times$ 0.3~ms and is far more computer-time intensive.

\begin{figure}[h]
\centering
\includegraphics[width=0.95\textwidth]{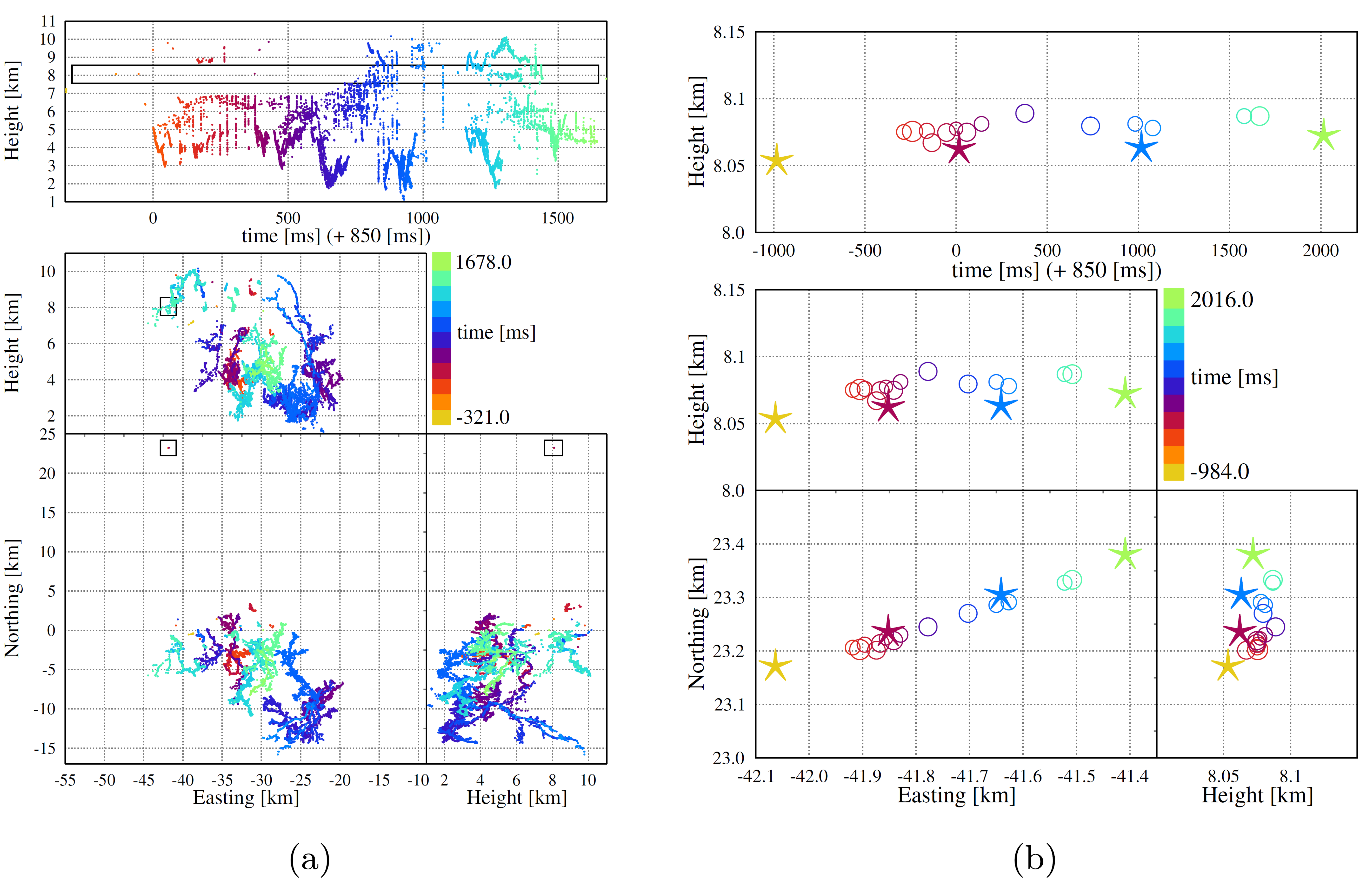}
	\caption{Side (a) provides an overview of LOFAR recorded lightning from flash 19A-1. The little specs at (N,E)=(23.2,-41.8)~km, marked by the square box, are the recorded events coming from the airplane. On side (b) the open circles give a zoom-in of the sources in this little spec, where the stars show the positions of the airplane in 1 second intervals as determined from flight data, and discussed in \secref{Results}. }
	\figlab{19A-1}
\end{figure}

For this work we focus on a single LOFAR recording, 19A-1 (at 2019-04-24, 19:44:32 UTC), that was also used in Refs.~\cite{Hare:2021, Hare:2023}. We have re-calibrated the 195 dual-polarized antennas distributed over 35 of the 38 Dutch LOFAR stations, about 6 antennas per station, following the procedure discussed in~\cite{Scholten:2021-init} while including some sources that were associated with the airplane. The nearest station was at a distance of about 20~km from the plane, the farthest at about 100~km.  The complete recording, imaged with our impulsive imager~\cite{Scholten:2021-init}, is displayed in \figref{19A-1}, left side, which shows that the major discharge during this recording period started at t=0, with very few located sources before this time. One may notice in this figure a small spot at (N,E,h)=(23.2, -41.8, 8.1)~km which, when zoomed in, expands to the image shown at the right of  \figref{19A-1}.

From the right side of \figref{19A-1} it is clear that we have observed emission from an airplane flying in an east-north-east direction at a speed of about 800~km/h at an altitude of 8~km, slightly climbing. This is confirmed from flight data, see \secref{Results}.
This airplane track is used in subsequent TRI-D imager~\cite{Scholten:2021-init} runs to search for sources in a volume of ($180\times 160\times 170$)~m$^3$, divided into voxels by a ($31\times 41\times 15$) Cartesian grid, 
co-moving with the track. This procedure is applied along the track from $t=-324$ till $t=-45$~ms where t=0 corresponds to the start of a major lightning discharge, using multiple TRI-D imager runs, each covering 0.3~ms. Keeping only the sources that have a strength that is at least twice as large as the strongest background source,
we are left with close to 700 candidate sources. We selected this particular start time on the track since at earlier times there was much lightning discharge activity near the location of the main flash, preventing a detailed imaging of weak sources close to the airplane location. Times after t=-45~ms were not investigated with the TRI-D imager. The grid was chosen such as to cover a sufficiently large volume around the expected airplane location with a grid that was sparse enough to keep the computing time within acceptable limits. Partly because of the rather sparse grid, these sources are not accurately imaged.

\subsection{The windowing issue of TRI-D imager}\seclab{TRID}

As mentioned before, the TRI-D procedure operates by gridding a tesseract in space and time and performing a beamforming procedure for every voxel in the tesseract, see \protect{\appref{TRID}}. This procedure is very useful for the detailed imaging of a small section of a lightning structure with a minimum of assumptions on the structures of the time traces as measured in each antenna, it also poses some imaging artifacts as discussed in this section.

A common artifact of TRI-D is that a single point-source that emits for longer than TRI-D's fixed integration time ( typically 100 ns) will be reconstructed as multiple source locations with a scatter of 20~m, see for example Ref.~\cite{Scholten:2022}. While this is already a very competitive result in the field of lightning science, where the resolution is often measured in hundreds of meters, this is not good enough for locating the VHF-sources on an airplane.
This spread occurs because in the TRI-D imaging technique the time trace of the beam for each voxel, in a 3-D voxelated space, is sliced into regular segments, typically 100~ns, over which the intensity is integrated and for each slice, the voxel with the highest intensity is identified. We find that this slicing process is the main source of these imaging artifacts.

\begin{figure}[h]
	\centering{ \includegraphics[width=0.65\textwidth]{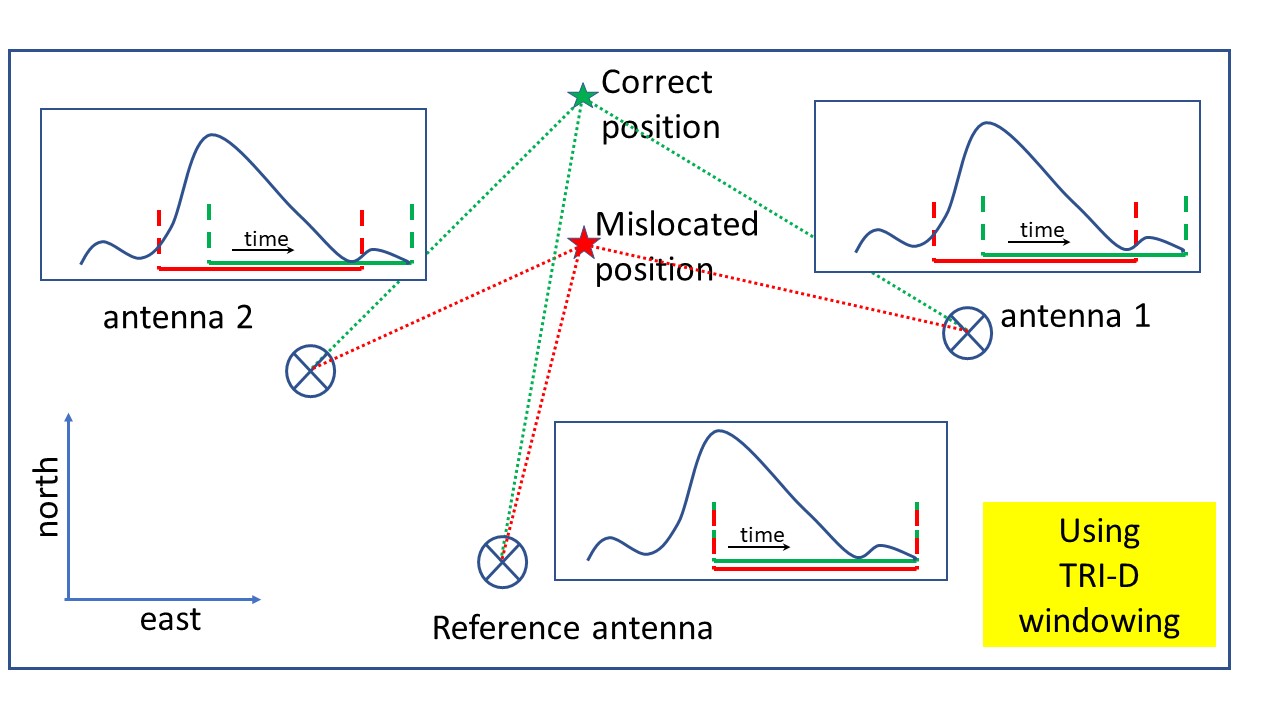}
	\includegraphics[width=0.65\textwidth]{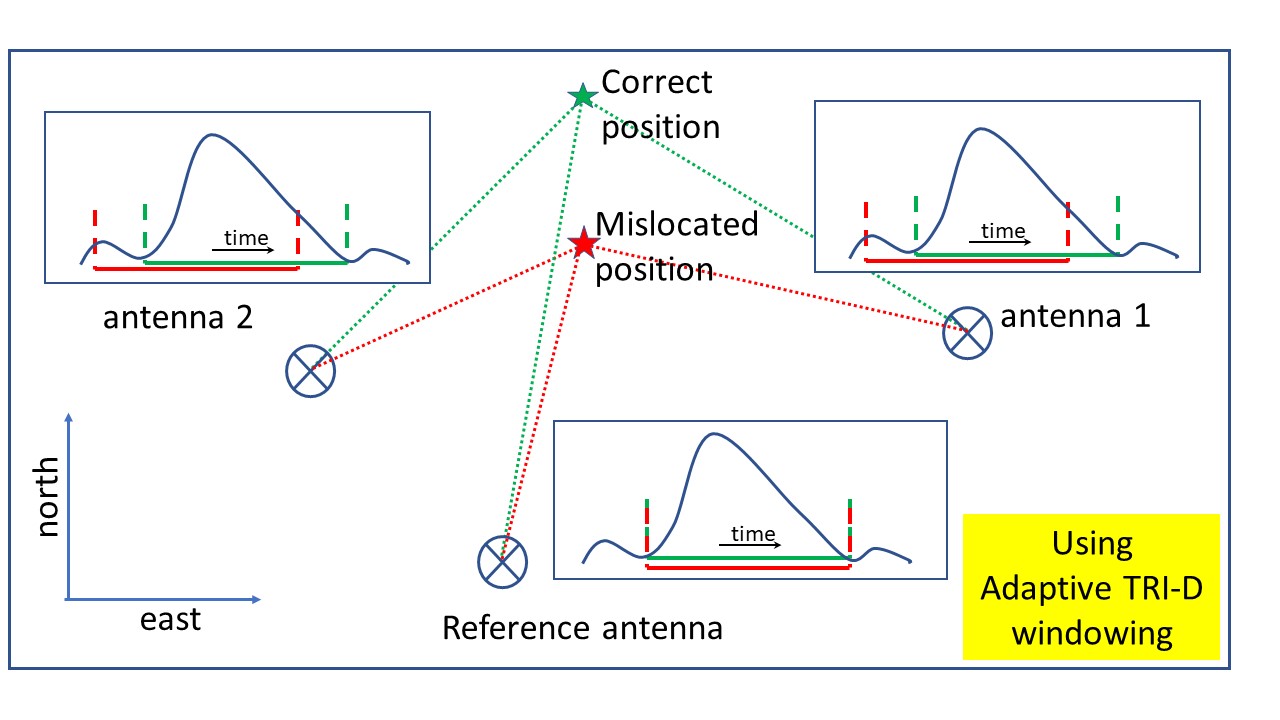}  }
	\caption{The two panels display a schematic map of the antenna layout to show the importance of properly adjusting the beamforming time window.
Next to the antenna locations, indicated by $\bigotimes$ symbols, the insets show a schematic time trace where, in green and red the time-windows are denoted for the TRI-D (top panel) and ATRI-D (bottom panel) procedures. In green (red) the window for the correct (mislocated) source positions are given, respectively. }
	\figlab{Windowing}
\end{figure}

Understanding this imaging artifact requires going into details of beamforming for broad-band pulses.
Since our pulses are very short we assume that they are emitted by a point source.
In finding the location of a such a point source through beamforming one searches for a location such that the signal amplitudes of all antennas, when correcting for the travel time of the signal, have the same phase. Adding these is a coherent process where the power of the summed signal reaches a maximum at the correct position of the source. For any different location the travel-time correction will be such that not the signal amplitudes of all antennas will add coherently, resulting in a lower power for the summed signal.
However, a different situation may arise when the intensity is calculated not for the complete pulse, but only for a part that falls within a pre-defined time-window, as is schematically shown in \protect{\figref{Windowing}} where the two panels display a schematic map of the antenna layout to show the importance of properly adjusting the beamforming window. The windowing times are always fixed for the reference antenna, usually one in the LOFAR core, while the position of the window for the other antennas will depend on the relative travel times of a signal from the assumed source position. In the TRI-D procedure the window is set arbitrarily and might thus cover only part of the pulse as is schematically shown in the top panel, while in the improved ATRI-D procedure (discussed in \protect{\secref{ATRID}}) it is set to cover the complete pulse (bottom panel). When performing beamforming at the correct source position (green in both panels) the time window will select the same part of the pulse as for the reference antenna. For the indicated mislocated position (red in both panels) the window will move to (relatively) earlier times for the other antennas. The beam-formed intensity, i.e., adding the pulses of all antennas within the window, will for the TRI-D imager (top panel) be larger for the red windows than for the green while for the ATRI-D procedure the green window gives the largest intensity.
For the example shown, mislocating the source closer to the core antenna, thus reducing the calculated travel time, moves the window to earlier times for the remote antennas for which the signal travel time remains the same. Thus, selecting a false source location where the time shifts due to path-length differences will increase the coherent-signal power in the pre-defined window.
Searching for the maximum intensity in the beam-formed signal will thus lead to wrong location determination for those cases in which the window is not properly centered. Note that the pulse shown in \figref{Windowing} is a simplification where the oscillatory nature of the signal is ignored, for simplicity sake.

\subsection{Improved Accuracy TRI-D imager}\seclab{ATRID}

The windowing issue is addressed in our adaptive TRI-D (\ATRID\ for short) imaging procedure by selecting the slicing window in such a way that it captures the entire peak of the pulse, or, alternatively, by placing the sides of the window at points where the intensity of the beam-formed time trace reaches a minimum, as is shown on the bottom panel of \figref{Windowing}. This adaptive windowing ensures that the optimal intensity for each source is minimally affected by parts of the pulse shifting in and out of the window (used for calculating the intensity) as the search for the maximum intensity voxel is conducted.
To perform this adaptive windowing procedure, especially for weaker sources where the pulse is barely visible in the time traces from antennas at the core, we construct the TRI-D beam-formed time trace at the initially suggested source location, where the beamformed time trace is used in the following step in the imaging procedure to define the adapted window.

The implemented procedure for constructing the window works as follows and starts from initial suggestions for the source location and time window.
Within the initially suggested window we identify the time sample, $t_{max}$, with the highest intensity in the beam-formed time trace for the originally set location. Then, we search for earlier ($t_e$) and later ($t_l$) times where intensity minima are found, either below a background level or, if this is not found, as the absolute minimum within twice the original window. The new window is then defined to run from $t_e$ to $t_l$. This method ensures minimal intensity at the borders of the slicing window, which is crucial for accurately determining the source position as explained by \figref{Windowing}. The procedure may result in a narrower window if the time trace has minima below the background within the originally set window, or enlarge the window.
For stronger pulses, the resulting adapted window typically spans around 100 time samples, or 500~ns, while for weaker pulses usually around 200~ns or less.

The thus selected window is used in the subsequent stage, using the beamforming procedure implemented in the TRI-D imager. As the originally set location for the source is not necessarily accurate, the volume used for the TRI-D run must be sufficiently large to encompass the true source. To minimize the number of voxels, and thus computing time, we employ a relatively coarse grid with a grid fineness of $F=2.3$ (see \appref{F}), which is still fine enough to identify the correct maximum.
For the present case this implies a volume of $140\times 140 \times 140$~m$^3$, with a voxel size of $3.4\times 2.4 \times 6.4$~m$^3$. These dimensions depend on the specific LOFAR antenna layout surrounding the area of interest.
In a following step, to improve localization accuracy, a second TRI-D run is performed, centered on the newly identified source location, using a finer grid with $F=0.75$ and a cubic volume of 30~m on each side. For the final source location, we apply the standard grid-interpolation procedure implemented in the TRI-D imager.
Such a two step approach to reach improved accuracy in an interferometry-based approach was first described by~\cite{Shao:2020} and named 2-step focused interferometry in \cite{Pu:2024}.

At this new location, we compute the fit quality, $Q$,  which indicates how well the time traces from individual antennas fit a time-dependent point source emitting dipole radiation positioned at the found source position.  The polarization density for the source, see \secref{Polarization} is derived by averaging the properties of this point source over the time window.

The fit quality indicator for a source, $Q$, behaves similarly to a $\chi^2$ value, see \eqref{Qual}, and is defined as such in Eq.~(3) in Ref.~\cite{Scholten:2022}, for fitting point sources to the antenna traces, but with an approximate normalization procedure due to the fact that the background (error bar) is difficult to estimate.
The reason is that not all time samples in the antenna traces are independent, owing to the effects of filtering with the antenna function, and that estimating the error bars or the amplitude accuracy is complicated by the presence of pulses from other sources during the flash. Despite these limitations, $Q$ remains a useful measure of the relative quality of a source as reflected in location accuracy.

In summary, the implemented procedure for the \ATRID\ imager starts with a reasonably accurate guess for the source location and consists of the following steps
\begin{enumerate}
\item {\bf Windowing;} Use the beam-formed time trace at the initial position to construct the correct imaging window, i.e., $t_e$ \& $t_l$.
\item {\bf Coarse TRI-D;} Run the TRI-D imager, centered at the guessed position, over a volume that is large enough to contain the true source position with a grid that is sufficiently fine to capture the intensity maximum.
\item {\bf Fine TRI-D;} Run the TRI-D imager, centered at the newly-found position, using a sufficiently fine grid to be able to find an accurate source position.
\item {\bf Quality \& Polarization;} Compare the time traces in all antennas with that of a modeled time-dependent source source at the source position and determine the source quality, $Q$.
\end{enumerate}
This procedure may be iterated a few times to achieve a stable result. Sources that have a quality $Q$ worse than a certain level should be removed.

In the \ATRID\ imaging procedure we locate the sources by maximizing the beamforming intensity by using the beamforming procedure implemented in the TRI-D imager. Alternatively one may envision a procedure focussed instead on optimizing the quality indicator $Q$. Such a procedure turns out to have some drawbacks; a) It is computationally more intensive than a beamforming intensity calculation. b) A steepest descent method for optimizing $Q$ has the difficulty that the $\chi^2$ surface has a complicated structure, reminiscent to that of the intensity surface. c) The location that gives the optimum value for $Q$ coincides with that for intensity for almost all cases we have investigated, and it is not clear which is better. We thus have selected the computationally simplest method with only at the very end a calculation of $Q$ to provide an independent selection criterium.

The time window used in beamforming is determined only at the start of the refinement procedure. Even though in some cases the location may have shifted by as much as 100~m, the structure of the beam-formed time trace changes little and there is little gain is optimizing the slicing window more frequently. By repeating the procedure once or twice a consistent result can be obtained.

\section{Results}\seclab{Results}

\subsection{The flight path}\seclab{f-path}

\begin{figure}[h]
\centering	
\includegraphics[width=0.95\textwidth]{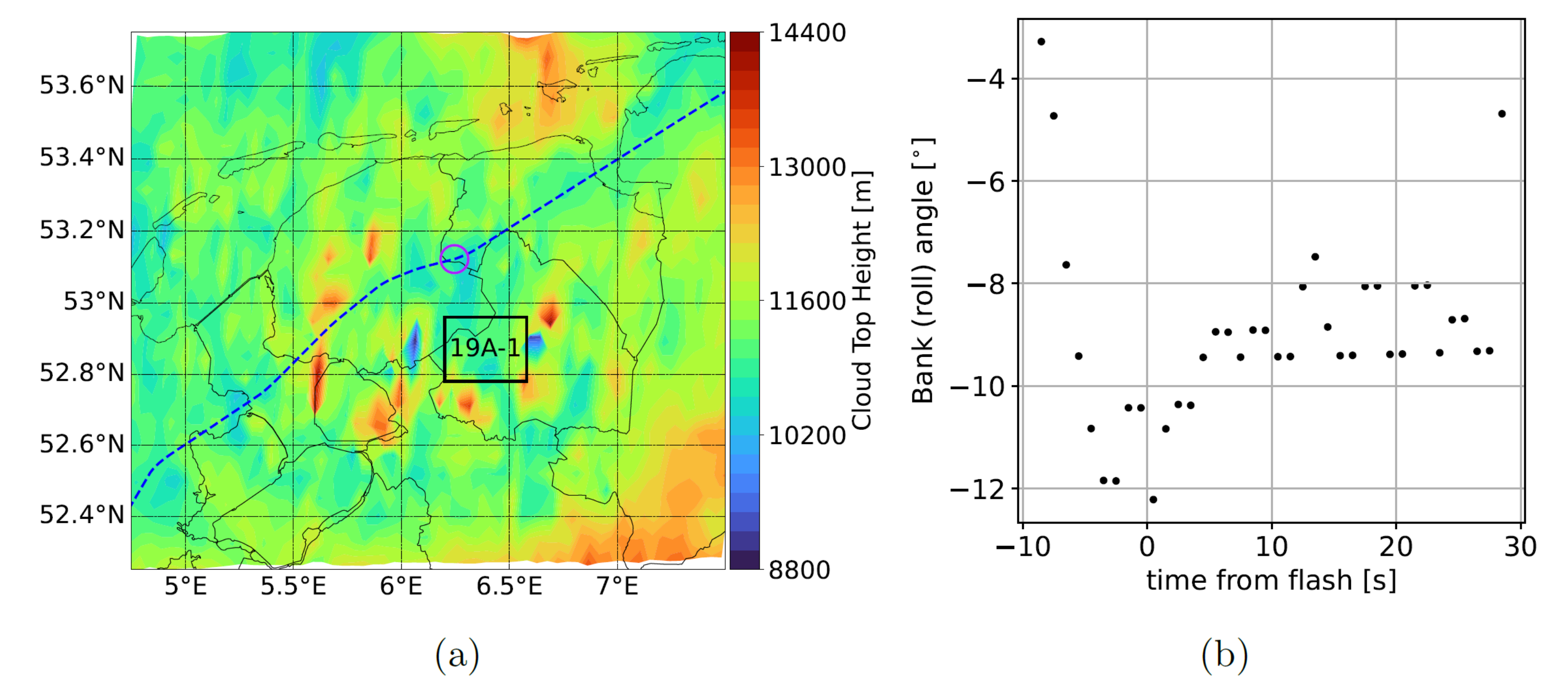}
	\caption{(a) The flight path of the plane based on historical flight data from The OpenSky Network~cite(Schaefer:2014) where the circle indicates the section of the flight path observed by LOFAR (see \figref{19A-1}). The background in this figure shows the parallax-corrected cloud-top heights as measured by MSG~cite(Finkensieper:2016, Meirink:2022). (b) Banking angle of the aircraft as a function of time in seconds where $t=0$ corresponds to the start of the flash.
}
	\figlab{flight_path}
\end{figure}

Using historical flight data from The OpenSky Network~\cite{OpenSky, Schaefer:2014}, we determined that the airplane was a Boeing 777-306ER (B77W) flying as a KLM flight from Amsterdam Airport Schiphol (EHAM) to Taiwan Taoyuan International Airport (RCTP). The flight data stored by OpenSky Network~\cite{OpenSky} was broadcast by the aircraft through Automatic Dependent Surveillance-Broadcast (ADS-B) and includes latitude, longitude, altitude, velocity, and heading in one second time intervals. The initial part of the flight path is shown in \figref{flight_path}, where the open circle indicates the section where we detected the plane with LOFAR.
The flight path is projected over a plot of parallax-corrected cloud-top heights as measured by the Meteosat Second Generation (MSG) satellite~\cite{Finkensieper:2016, Meirink:2022}. Since the altitude of the plane was about 8~km, this indicates that the plane was flying through some high-altitude clouds.
Thus one would expect that the aircraft becomes electrically charged~\cite{Guerra-Garcia:2018, Martell:2022}.

According to the flight data, when LOFAR detected the aircraft, it was flying with a velocity of 220~m/s with a heading of roughly $73^{\circ}$ from magnetic North, still climbing after having taken-off from Schiphol. These quantities match the velocity and heading we can derive by analyzing the distribution of VHF pulse sources in space and time 
as shown at the right of \figref{19A-1}, where the flight path data from OpenSky Network together with the impulsive imager data are plotted in LOFAR coordinates for the duration of the lightning flash. LOFAR coordinates are given in a Cartesian system where the vertical direction is the direction of a plumb line at the LOFAR-core, the Superterp. Since the plane is observed at a distance of about 50~km from the core, the curvature of the Earth introduces a vertical off-set of roughly 220~m between the LOFAR coordinates and the GPS coordinates reported by the airplane, which has been corrected for. The right of \figref{19A-1} shows that the locations of the VHF sources imaged using the impulsive imager, line up with the GPS location of the aircraft, within 20~m vertically and about the same in the horizontal plane. This accuracy is getting at the same level as the size of the plane and it thus becomes important where precisely the GPS-tracker is located on the plane. We have re-adjusted the relative timing between the ADS-B and the LOFAR data by about 300~ms, well within the accuracy of ADS-B timing.

The flight data does not provide the bank angle directly. To obtain the results shown in the right panel of \figref{flight_path} we assumed that the airplane was not climbing and used its velocity and the time derivative of its heading direction, $\psi$,
\beq
R_t \approx {v} \left(\frac{\mathrm{d} \psi}{\mathrm{d} t}\right)^{-1} \;,
\eeq
to recover the turning radius as a function of time. The banking angle was computed using equation 6 from Ref.~\cite{Sun:2019} setting $\gamma = 0$. From this figure it is seen that the plane was reaching a maximal banking angle of about -12$^\circ$ just at the time of the LOFAR observation, where a negative angle implies that the left wing is lower than the right.

\subsection{Evaluation of the \ATRID\ imager accuracy}

\begin{figure}[h]
\centering{%
\includegraphics[bb=0.5cm 2cm 208cm 128cm,clip, width=0.98\textwidth]{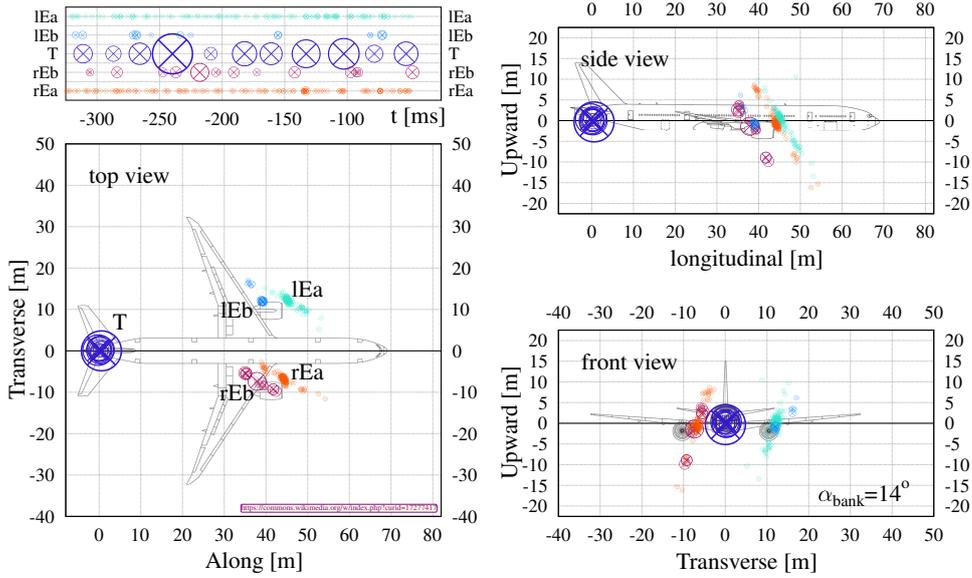}%
}
	\caption{The imaged VHF-sources transformed into a frame that is co-moving with the plane. The `Along'-axis is pointing in the flight direction of the plane. A bank angle of -14$^\circ$ has been accounted for.
The size of the wagon wheels indicates intensity of each source.
The color coding of the events is following the position on the airplane where `rEa', `rEb', `T', `lEb', `lEa' label right engine front, right engine back, tail, left engine back, and left engine front respectively.
Only analyzed data between t=-325 and -45~ms is shown in the figure. The drawing of the airplane, a Boeing 777-300ER is taken from Ref.~cite(Boeing777-300ER).}
\figlab{Airplane}
\end{figure}

The results on locating the sources linked with the airplane, using \ATRID, as detailed in \secref{ATRID}, are shown in \figref{Airplane} in a reference system that is co-moving with the airplane (as determined from the ADS-B data) accounting for a -14$^\circ$ bank angle.
While the flying direction is taken from the flight data, the bank angle of -14$^{\circ}$ (slightly larger than the angle estimated from \figref{flight_path}) is obtained by aligning the source positions in the transverse direction (transverse to the flight direction) with the image of the plane, a Boeing 777-300ER is taken from Ref.~\cite{Boeing777-300ER}, where increasing the bank angle will lower the side with negative transverse distance (right) with respect to the positive (left) side.
From the right side of \figref{flight_path} it appears that the plane is just entering the turning motion where initially the bank angle shows an overshoot reaching close to -12$^{\circ}$ at the time of the LOFAR observation before entering in the stable bank angle of -9$^{\circ}$ while continuing the course-correction manoeuvre. It might be that the actual bank angle is somewhat different from our relatively simple estimates because, for example, effects of wind have been ignored.

From \figref{Airplane} it is clear that we observe sources associated with five different locations on the plane. We have labeled them as tail (T) sources and as left and right engine (lE, rE respectively) where for each engine we observe sources more related to the front (a) and the back side (b), thus lEa, lEb, rEa, rEb.
The positioning of the picture of the plane (a Boeing 777-300ER) in \figref{Airplane} with respect to the detected sources is adjusted to put the Ea sources at the front- and the Eb sources at the back-ends of the two engines since the relative spacing of these sources match the engine size. This puts the tail pulses some 4~m in front of the tail-end of the airplane. However, alternatively, one could position the tail sources at the tail-end of the plane, putting the Ea sources at the middle of the two engines and the Eb sources somewhere in the middle of the wing behind the engines.

The size of the wagon wheels in \figref{Airplane} is an indicator of the intensity of each detected source. Clearly the ones near the tail are the most powerful ones. The panel at the upper left shows the time dependence of the events associated with the different positions on the plane. Each horizontal row in this panel corresponds to a different source location. The central row shows the events associated with the tail. The average time between two tail events is about 25~ms but varies between 20 to 30~ms. The quasi-regularity of these events could be explained by a constant charging of the plane due to friction with ice crystals in the cloud~\cite{Martell:2022}. It could also be caused by some quick loading or unloading of a capacitor, possibly associated with the auxiliary power unit (APU) positioned in the tail of the plane. Signals of malfunctioning electronics on a plane have been recorded before by LOFAR~\cite{Wucknitz:2011}. There have since been several other sightings of airplane tracks by radio observatories~\cite{Aab:2016, RNO-G:2025-plane}.  See \secref{Physics} for a more extensive discussion.

\begin{table}
 \begin{tabular}{|l|r|ccc|ccc|}
  \hline
   & \# & \multicolumn{3}{c|}{PCA standard dev.\ [m]} & \multicolumn{3}{c|}{airplane fixed} \\
  Category &  & $\sigma_1$ & $\sigma_2$ & $\sigma_3$ & along & trans & up \\
  \hline                                              
right Ea & 115 &  4.75 & 0.61 & 0.28 &  0.48 & -0.27 & -0.84 \\
right Eb   &  16 &  5.21 & 0.66 & 0.20 &  0.47 & -0.27 & -0.84 \\
tail         &  11 &  0.57 & 0.29 & 0.19 &  0.09 &  0.39 & -0.92 \\
left Eb    &  14 &  3.50 & 0.45 & 0.15 &  0.43 & -0.60 & -0.68 \\
left Ea  & 105 &  4.15 & 1.11 & 0.37 &  0.46 & -0.28 & -0.84 \\
   \hline
 \end{tabular}
\caption{A principal component analysis (PCA) of the covariance matrix for source locations in airplane-fixed coordinates. The second column shows the number of sources that have been associated with each location on the plane (first column). The following three columns show the standard deviation ($\sigma$, in units of m) for the three PCA axes and the last three columns give the orientation of the main axis in airplane-fixed coordinates.}\tablab{LocPCA}
\end{table}

To quantify the source location accuracy we present in \tabref{LocPCA} the results of a principal component analysis (PCA) for the sources in each of the five categories in airplane-fixed coordinates. Of the in total 278 sources that have passed the quality selection criterion $Q<1.1$, 17 have been excluded from this analysis since they were too far from the plane (more than 15~m). \tabref{LocPCA} shows that the similarities between the distribution of the different groups of engine events is quite striking. They have similar spread along the main error-axis as well as in the minor directions where also the orientation of the main axis is very similar. The distribution of the tail events is, however, rather different, the spread along the main axis is only about 60~cm, mostly vertically distributed, a very different orientation from that of the other sources. Even for the two categories with over a hundred sources, the source distribution is far from spherical with a long axis of the order of 4~m and a short axis of order 0.3~m and very similar to that of the Eb events with a much lower count and a much larger strength (see \figref{Airplane} and \tabref{AvePol}). This result suggests that the intrinsic resolution might be as good as 30~cm, almost independent of source strength, even for sources that can barely be distinguished from background for the core antennas.
This compares well with the 10~cm determined from simulations in Ref.~\cite{Scholten:2022}.
Analyzing the direction of the long axis shows that in Earth-centric coordinates it is oriented about 45$^\circ$ down and in the direction of the LOFAR core. A possible interpretation of the observed spread in source locations is discussed in \secref{Physics}.  The fact that this spread is roughly oriented along the line-of-sight to the LOFAR core, suggest that it is instrumental.

\subsection{Polarization analysis}\seclab{Polarization}

\begin{figure}[h]
\centering{%
\includegraphics[bb=0.5cm 2cm 208cm 128cm,clip, width=0.98\textwidth]{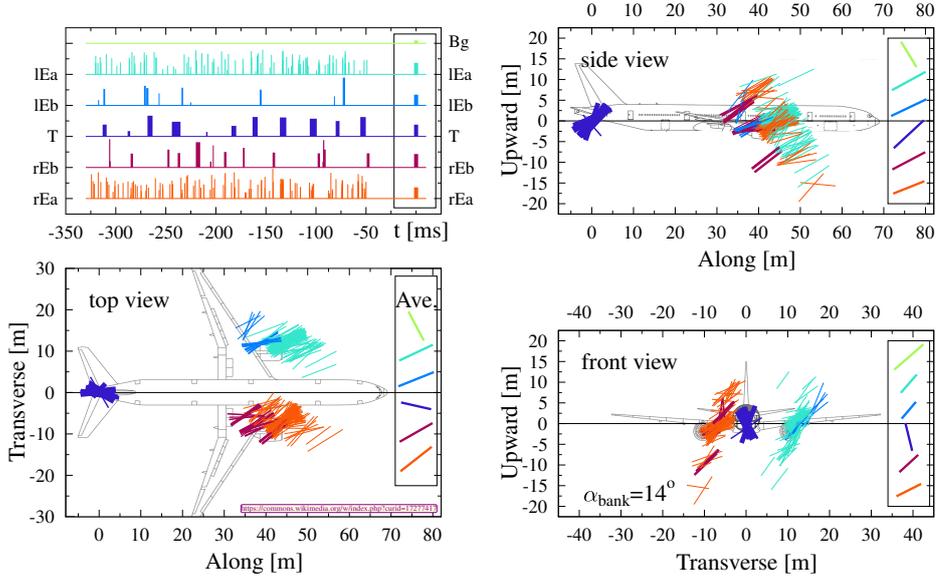}%
}
	\caption{The analysis of the main linear polarization direction of each source as determined through the principal component analysis discussed in \appref{pol}. The color coding of the events is following the position on the airplane where `rEa', `rEb', `T', `lEb', `lEa' label right engine front, right engine back, tail, left engine back, and left engine front respectively. The widths of the bars is proportional to the square root of the strength of the principal linear polarization component for all panels.
In the top left panel the length of the bars indicate the fraction of linear polarization.
The inset on the right of this panel shows the same for the average of the events (width of the bar is constant), the same information as is given in \tabref{AvePol}. The inset shows also (green color, labeled as Bg) the equivalent value for background sources (see text). In the bottom left panel and right panels show the projection of the (normalized) 3-D polarization orientation of each source on different planes that are aligned with the airplane.
}
\figlab{AirplanePol}
\end{figure}

The TRI-D imager does not only provide an accurate location of each source, but, since each source is modeled as a (time-dependent) point source, also its 3D polarization density, integrating the polarization direction over the time-window used in the beamforming algorithm. The polarization density is analyzed using a principal component analysis as discussed in detail in \appref{pol}, where the results for the main linear polarization component are shown in \figref{AirplanePol}. Each source is presented by a bar of unit length and a width that is related to the strength of the main linear polarization component. The projection of this bar on the different airplane-centered planes is what is shown in \figref{AirplanePol}. In the top left panel of this figure the fraction of linear polarization is shown. The insets on the right side of each panel show the same quantity as the panel itself, only averaged over all sources of a particular category; rEa, rEb, tail, lEb, and lEa. Also shown, for comparison, is the quantity for a 0.3~ms section of time where no airplane or other sources are distinguished, and is labeled as background (the data on separate background sources is not shown). The width of these summary bars is just constant. To obtain the averaged quantities we sum the polarization densities of all sources for a specific category and subsequently apply the same principal component analysis as for a single source.

\figref{AirplanePol} shows that, surprisingly, the weaker sources, those associated with the engines of the airplane, show as much scatter in polarization direction as the stronger ones.  In a following section we will have a closer look at the structure of some of these sources. One also notices that all engine-associated events for both sides of the plane are very similar; the polarization directions are basically identical and the fraction of linear polarization is about 0.4 (see also \tabref{AvePol}). The tail events also have a substantial linear polarization component but with a more vertical orientation. In a TRI-D analysis the background is known to carry a net polarization which is most probably due to imaging artifacts related to the very in-homogenous distribution of the antennas around the location of interest~\cite{Wang:2024, Turekova:2024}. The polarization direction and degree of linear polarization of airplane sources is clearly different from background sources as can be seen from \tabref{AvePol}where $\theta$ is polar angle, with $\theta=0$ upward, and $\phi$ is the counter-clockwise azimuth angle where $\phi=0$ is along the aircraft. Note that for a polarization directions, ($\theta$, $\phi$) is indistinguishable from ($180-\theta$, $\phi-180$).

\begin{table}
 \begin{tabular}{|l|c|ccc|cc|}
  \hline
  Category & $\bar{W}$ & $P_{\rm un}$ & $P_{\rm lin}$ & $P_{\rm circ}$ & $\theta_{\rm lin}$ & $\phi_{\rm lin}$ \\
  \hline
  right Ea &  1.4 &  63 &   36 &    0.9 &         73 &   38  \\
  right Eb  & 14. &  55 &   43 &    2.0 &         65 &   30  \\
  tail        & 76. &  61 &   38 &    1.1 &         48 &  -13  \\
  left Eb   &  6.3 &  63 &   34 &    3.1 &         66 &   22  \\
  left Ea &  0.9 &  62 &   37 &    0.6 &         64 &   25  \\
  background  &  .11 &  91 &    9 &    0.1 &         52 &  118  \\
  \hline
 \end{tabular}
\caption{Given are, per category of the sources,
the average power, $\bar{W}$ in units of [mW/MHz] at 60~MHz,
and the average polarization probabilities (in \%).  Also the average linear polarization angles (in degrees) are given where $\theta$ is polar angle, with $\theta=0$ upward, and $\phi$ is the counter-clockwise azimuth angle where $\phi=0$ is along the aircraft. For a polarization directions, ($\theta$,$\phi$) is indistinguishable from ($180-\theta$,$\phi-180$). }\tablab{AvePol}
\end{table}

As can be read from \tabref{AvePol} the percentage of circular polarization is small. A likely source for circular polarization is two sub-sources that have different linear-polarization directions where one occurs very shortly after the other, such that they overlap in time~\cite{Scholten:2016}. Since the order in which these occur is probably random, one expects that the average circular polarization for many sources will be vanishingly small, as is indeed the case.

For all categories the percentage of linear polarization for the average is about 10 percentage points smaller than the mean linear polarization for each source (not shown in \tabref{AvePol}). Since this is due to the scatter in the main polarization direction for each category, we conclude that the spread in linear polarization angle is about 25$^\circ$. The precision of our analysis will probably be better than this, since part of this will be due to genuine scatter in the polarization direction of the sources.

The column labeled as $\bar{W}$ in \protect{\tabref{AvePol}} gives the emitted power of the pulse, i.e., its energy divided by the duration, per MHz using the procedure discussed in Ref.~\protect{\cite{Scholten:2023PL}}. We have opted for units of [mW/MHz] at 60~MHz since background can only be discussed in terms of power and because the the power is measured in a limited frequency range around 60~MHz, where the antennas have a peak sensitivity. The power of the tail sources are of a comparable magnitude as a medium-power pulse seen in lightning, however the latter cover several orders of magnitude~\protect{\cite{Machado:2021}}.

\subsection{Time traces of ``tail" sources}\seclab{TimeTrace}

\begin{figure}[h]
\centering
\includegraphics[width=0.98\textwidth]{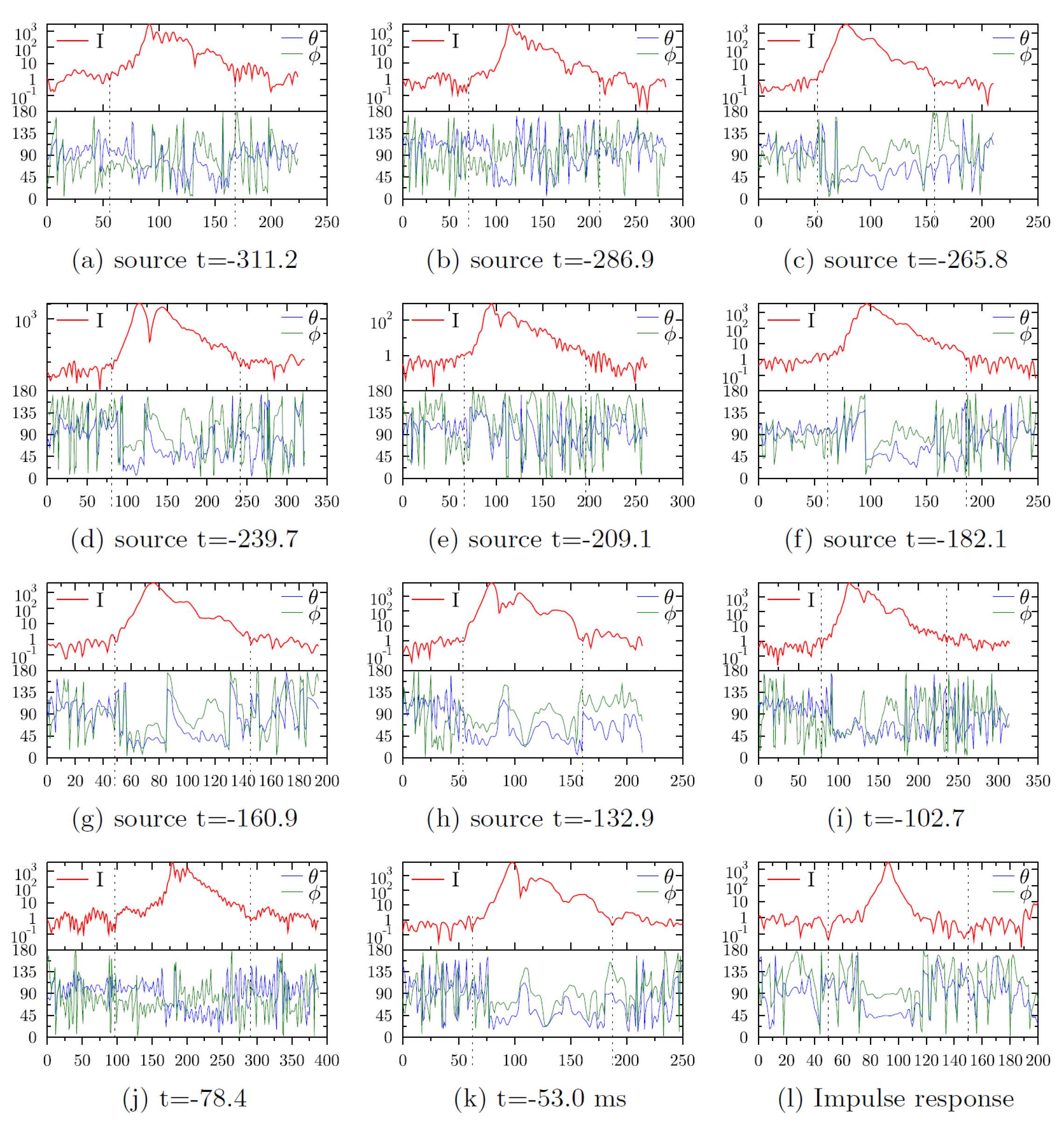}
	\caption{Each panel details the coherent time trace (I) and the direction ($\theta$, $\phi$, in degrees, see \tabref{AvePol}) of linear polarization for a source localized near the tail of the aircraft as function of time sample (of 5~ns).  The last panel shows the impulse response of the system. The thin vertical dashed lines at 1/4 and 3/4 of the abscissa indicate the window, i.e., $t_e$ and $t_l$, as used in the analysis.}\figlab{TailPulses}
\end{figure}

\begin{table}
 \begin{tabular}{|l|c|cc|ccc|cc|}
  \hline
    label &  t [ms] & ${I}$ & Width & $P_{un}$ & $P_{lin}$&$P_{circ}$& $\theta$& $\phi$  \\
  \hline
   a &  -311.15   &  2.4 &  112 &   60.7 &   38.1 &    1.1 &   66&  75   \\
   b &  -286.90   &  2.1 &  141 &   78.5 &   17.3 &    4.2 &   37& 102   \\
   c &  -265.77   &  3.1 &  105 &   33.1 &   66.8 &    0.1 &   39&  71   \\
   d &  -239.66   & 15.3 &  161 &   46.4 &   47.3 &    6.3 &   40&  91   \\
   e &  -209.06   &  1.4 &  131 &   52.1 &   14.6 &   33.2 &   72& -44   \\
   f &  -182.08   &  4.9 &  124 &   61.9 &   34.0 &    4.1 &   37&  11   \\
   g &  -160.85   &  2.9 &   97 &   36.4 &   62.7 &    0.8 &   37&  64   \\
   h &  -132.873  &  5.3 &  107 &   36.0 &   61.0 &    2.9 &   40&  70   \\
\hdashline
  h1 &  -132.8732 &  3.5 &   32 &   10.0 &   87.3 &   2.71 &   41&  70   \\
  h2 &  -132.8729 &  1.8 &   75 &   60.3 &   35.0 &   4.71 &   34&  64   \\
\hdashline
   i &  -102.72   &  9.3 &  157 &   41.3 &   57.3 &   1.44 &   59&  59   \\
  j &   -78.40    &  3.6 &  194 &   49.6 &   49.7 &   0.71 &   61&  75   \\
  k &   -52.971   &  4.4 &  125 &   35.4 &   62.5 &   2.18 &   41&  72   \\
\hdashline
 k1 &   -52.9711  &  1.0 &   43 &    9.5 &   88.5 &   1.95 &   41&  70   \\
 k2 &   -52.9709  &  3.2 &   43 &   62.1 &   32.1 &   5.82 &   35&  67   \\
  \hline
 \end{tabular}
\caption{The complete list of identified tail sources. The label corresponds to the sub-figure label in \figref{TailPulses}. The quantities shown are equivalent to those given in \tabref{AvePol} with the exception that the column ${I}$ gives the total (pulse integrated) power in units of [nJ/MHz] at 60~MHz.
For pulses h and k also the results for the earlier and later parts are given. The pulse width is given in units of samples (5~ns). }\tablab{TailPol}
\end{table}

The sources that are localized near the tail section of the aircraft are detailed in \figref{TailPulses}, where each sub-figure consists of two panes, the top showing, per time-sample of 5~ns, the intensity of the coherent time trace (in arbitrary units) and the bottom showing the linear polarization angle (in degrees). The thin dotted lines indicate the time window used for analyzing the properties of the source. It can be seen that the window start and stop times, named $t_e$ and $t_l$ in \secref{ATRID}, have been chosen suitably.
The abscissa in each panel marks time samples. The circular polarization per time sample is not shown, partly because this tends to average to almost zero over the full time trace and partly because it is related to the change of the linear polarization angle and thus contains little additional information. The figure shows that, at first glance, all tail pulses are very similar, but differ in detail. This is substantiated by the more quantitative comparison presented in \tabref{TailPol}.

The common features, see \figref{TailPulses}, are an exponential rise with an e-folding rate of about 15~ns and a more gradual fall-off with an e-folding rate of about 60~ns, where often some sub-structure is visible. The initial rise is as one would expect from the impulse response of the system, see last sub-figure labeled ``impulse response", while the fall off is much more gradual.
This implies that the sources are formed by a strong initial VHF pulse of very short duration, considerably smaller than 15~ns, followed by some exponentially weaker ones. This picture is substantiated by the linear polarization angles of the pulses.
For most sources one can clearly distinguish 2 follow-up pulses after the main one where all three have a very similar linear polarization orientation.
For the pulses at t=-265.8, -182.1, -160.9, -102.7, and -78.4~ms one still can distinguish different sub pulses, however they change in orientation roughly in the middle of the source window.
Although the sources at t=-311.2, -286.9, and -209.1~ms have a very similar intensity structure as the others, the polarization observables are difficult to interpret. It might be that this structure is caused by sub-sources that are somewhat (order 1~m) displaced in space and interfere. 

\tabref{TailPol} shows that, integrated over the complete pulse, the circular polarization is vanishingly small for most events except for the relatively weak pulse labeled `e'. For pulses `h' and `k' the \ATRID\ imaging procedure could run stably on the sub-structures of these pulses, showing, see \tabref{TailPol}, that the parts have a very similar polarization direction as the total. It is interesting to observe that the stronger, initial, sub pulses have a large linear polarization component and the weaker ones have an unpolarized fraction of close to 60\%. This is supportive of the picture where each composite source starts off with a single very strong, impulsive, spark followed by a few weaker ones in somewhat different directions. The sub pulses of `h' are located about 0.5~m from each other while those for `k' are about 1~m apart (not shown in the table). The complete source is located in between that of the two sub-sources for the two cases.

\section{Physics interpretation}\seclab{Physics}

The sources located near the engines exhibited a distinctive distribution, with the largest spread, with a standard deviation of approximately 4 meters, in a downward direction at an angle of about 45$^\circ$ toward the LOFAR core. In contrast, the spread in perpendicular directions is significantly smaller, nearing 20~cm. To investigate a possible explanation for this, we conducted simulation calculations involving $N_d$ impulsive point sources, all having the same strength, distributed randomly within a cloud that followed a Gaussian density profile characterized by a $1/e$ distance of $\sigma_s$. The point sources were also temporally spread with $\sigma_{time}=50$~ns. This is akin to the simulation methodology employed in Ref.~\cite{Scholten:2022}.
Although we have not exhaustively examined the various parameters in this simplified model, we obtained an interesting result. In one such simulation we have made 20 realizations of such a cloud of point sources, where each cloud consists of $N_d$=10 point sources distributed with $\sigma_s$=2 m. Since these sources emit in in a very small timespan, their emissions will interfere where the interference varies with the line of sight. These line of sight variations depend on the particular Mont-Carlo realization of the source cluster and and almost vanishes for for a large number of sources.
In the next step we used ATRI-D to determine  the location for each of these clouds. Analyzing these positions using a principal component analysis, very similar to what was used in obtaining \protect{\tabref{LocPCA}}, we found a distribution density with a long axis of 3.5~m and a short axis of 25~cm.
This distribution closely resembles that given in \protect{\tabref{LocPCA}} for the sources associated with the engines, including a similar orientation for the long axis. Increasing $N_d$ appears to reduce the spread, likely due to the central limit theorem. Therefore, the lack of a similar source distribution for the tail events may be seen as strong evidence that each detected source corresponds to  single point-like source, as opposed to the engine events, where each may correspond to a diffuse cloud of sources with an intrinsic spread of a few meters.

The engine events for the left and the right sides of the plane show very similar distributions, in density of pulses per time interval, in spread in location, in intensity and well as in polarization. It is thus intriguing to find that the distribution for the front and the back side of the engine are different in intensity as well as the number of pulses per time interval, while still showing very similar polarization direction and similar spread in location.
Since it is known that aircraft discharging may also occur through the engines~\protect{\cite{Jones:1990, Martell:2022}} we speculate that the VHF emission is due to the discharging of the plane although this could also be just from the running engines themselves. To sort this out requires a more comprehensive search for VHF emission from planes. The similarity in the distributions in space and polarization could indicate that the basic physics causing these sparks are very similar, where, since the propagation velocity of sparks (ranging from $10^5$ to $10^7$~m/s) is so much larger than the velocity of the plane (order 200~m/s), they escape as easily from the front of the engine as from the back. At the back the engine the hot air is more conducting which might cause larger but fewer sparks. It is interesting to note that at the back of the engines there are about ten times fewer pulses seen than at the front, which are about ten times stronger.
The fact that the plane is still climbing, with the engines thus working full throttle, may even enhance this effect.
This might also explain why the simulation, with point sources randomly distributed in a cloud with a diameter of 2~m (maybe not coincidentally the same as the diameter of the engines) gives rise to an localization distribution that is very similar to what we observe with the ATRI-D imager.
The small, but unmistakeable, offset of the engine events from the centers of the engines could be due to the fact that the tail events, being the brightest, were placed on the central axis of the plane. If these, in reality, were emitted at -1~m in the transverse direction, the picture of the plane would have to shift, placing the engine events at the centers of the two engines.

One possible scenario for the tail events is that the plane, flying through high clouds, through frictional charging with ice particles in the cloud~\cite{Martell:2022}, is constantly acquiring static electricity. At a certain time the charge on the plane exceeds a threshold where the electric field at certain points on the plane exceeds the breakdown value and a spontaneous discharge occurs. Since this involves strong changes in local current densities at short time scales, broad-band electromagnetic waves will be emitted, including in the VHF range of interest for this work. Since the charging will occur at a relatively steady rate, due to the constant friction with the ice particles in the cloud through which the plane is flying, the discharging will occur with a quasi-regular pulse as seen for the tail sources.
However, one expects this discharge to occur at protrusions from the plane, such as the wing-tips, the tail end of the plane, and mostly from one of the p-static wicks installed on the wings and tail of the plane for the purpose of releasing static electricity. Our observations show absolutely no evidence this is occurring. Additionally, one should expect a modulation in the engine events with the frequency of those at the tail if both were caused by electric charging of the plane.

Alternatively, from a physics point much less interesting, is that the short VHF radio pulses where the emission point is located at the tail section of the plane is emitted due to electronics inside the plane. At precisely the place of the tail from where we detect strong VHF emission, some four meter before the tail end, the auxiliary power unit is placed. At take-off of a plane, and often some time thereafter, this power unit is used to power an electrical generator as well as an air conditioning unit. It seems likely that any of these is responsible for the emission. The polarization component of the emitted radiation that is parallel to the local conducting surface of the plane will be absorbed and only the perpendicular polarization component will penetrate. This is in agreement with our observation of the polarization of the tail events.

We have also been able to observe the VHF source positions on another plane, an Airbus, flying at 11.7~km altitude in the close vicinity of the LOFAR core on August, 14, 2020 (recording 20A-8), where the results of a first analysis are presented in the supplementary information.  Also this plane flew through high clouds and we could detect emission near the tail as well as associated with the engines. Much like the case discussed extensively in this work, the tail events showed very little spread in contrast to the engine events and the intensity of the engine events was of the same order of magnitude. The tail events were also quasi-regular, however much weaker, of the same order as the engine events, and occurred with a much higher frequency of about 1 pulse per 2.4~ms. In the same LOFAR recording, ADS-B data showed  also a plane at about the same location as discussed in the main part of this work, flying at an altitude of 11.9~km, above the local cloud coverage. With LOFAR, we were not able to detect any VHF emission from this plane. Although we have not analyzed this case in as much detail as the one discussed in this work, it gave us supporting evidence for the claim that the engine events are a signature of electric discharging of the plane, while origin for the tail events probably lies in the electronics of the plane.

\section{Conclusions}\seclab{Conclusions}

The emission from the airplane appears rather well localized to a few specific locations on the airplane.  This gave us a unique chance to determine the accuracy of our observation technique under field conditions. Because of these well localized sources we had the opportunity to improve our imaging method.
Most important in this respect was the observation that most of the imaging artifacts could be avoided by adjusting the time window, used in the TRI-D beamforming analysis, to cover the complete extent of the source duration.
This allowed us to determine the location of the events near the tail of the plane with an accuracy of about 50~cm. The sources associated with the two engines show a very asymmetric spread. Evidence is presented that this spread could be due to the fact that the observed pulse is due to the emission from several point sources spread over a cloud of a few meters diameter.
It is thus, unfortunately, difficult to make very definitive statements on the intrinsic resolution on the LOFAR lightning imaging other than that it is definitely better than 50~cm, and probably even better than the 30~cm we observe for the length of the minimal axis in the PCA analysis of more than 100, relatively weak, sources associated with the front ends of each engines.

Improving the imaging technique has the direct spin-off that we can greatly improve the sharpness of the images made from lightning discharges. In particular, the proven ability to extract the polarization direction of the pulses will greatly help to improve our insight in the processes that generate lighting-pulses.

Analyzing the polarization of the different events we observe that those associated with the tail have a distinctly different direction from those associated with the engines. The tail sources have a dominant orientation pointing almost perpendicular to the plane. This is seen as evidence that the polarization direction of strong sources can be reconstructed correctly.

Assuming that the short VHF-radio pulses observed by LOFAR are due to electrical discharges caused by charging of the aircraft, the source locations and polarization directions seen in \figref{Airplane} and \figref{AirplanePol} are difficult to explain, showing that more work is needed to understand how electrical discharges are produced by aircraft and how these discharges generate VHF emissions.  Since lightning strikes to commercial aircraft often initiate as electrical discharges from the aircraft~\cite{Guerra-Garcia:2018}, such work could potentially benefit lightning protection as well.

To summarize, in this work we reported on a serendipitious observation of an airplane while observing a lightning discharge with the LOFAR radio telescope.
Proving that it is possible to observe localized short radio pulses emitted from an airplane, while flying at an altitude of about 8~km through high clouds, has several interesting applications. From a purely academic perspective it allowed us to improve our detection technique and clearly prove the accuracy in position and polarization direction of LOFAR observations of near-field sources of short radio pulses. Additionally, the observation of large electric sparks coming off such an airplane would pose much tighter constraints on modeling and the understanding of the electric currents in lightning discharges. Such sparks are known to occur, but, most unfortunately for us, did not happen during the present observations and we only observed short sparks. Related is that observations like ours test the action of static wicks under natural operating conditions. To our great surprise we found that we did not detect any emission from these static wicks. To perform a more comprehensive study of airplane discharging will require further dedicated observations with LOFAR of planes flying through high clouds. We have shown that this is a realistic endeavor.

\section*{Availability of data and materials}

The data are available from the LOFAR Long Term Archive (for access see \cite{LOFAR-LTA}), under
the following locations:
\\ \noindent {\footnotesize \verb!L703974_D20190424T194432.504Z_stat_R000_tbb.h5! }
\\ \noindent all of them with the same prefix
\\ \noindent {\footnotesize \protect{\verb!srm://srm.grid.sara.nl/pnfs/grid.sara.nl/data/lofar/ops/TBB/lightning/!} }
\\ \noindent
and where
``stat'' should be replaced by the name of the station, CS001, CS002, CS003, CS004, CS005, CS006, CS007, CS011, CS013, CS017, CS021, CS024, CS026, CS030, CS031, CS032, CS101, CS103, RS106, CS201, RS205, RS208, RS210, CS301, CS302, RS305, RS306, RS307, RS310, CS401, RS406, RS407, RS409, CS501, RS503, or RS508.
\\ \noindent
To access this data, please create an account following instructions at \protect{\cite{LOFAR-LTA}} and follow the instructions for ``Staging Transient Buffer Board data''. In particular the utility ``wget'' should be used as in
\\ \noindent {\footnotesize
\verb!wget https://lofar-download.grid.surfsara.nl/lofigrid/SRMFifoGet.py?surl=location! }
\\ \noindent where ``location'' is the location specified in the above.
\\The software used for data analysis is available at \cite{Scholten:2025-LOFLI-v24}.
\\\figref{19A-1}, \figref{Airplane}, \figref{AirplanePol}, and \figref{TailPulses} in this work have been made using the Graphics Layout Engine (GLE)~\cite{GLE} plotting package. The data displayed in these figures may be retrieved from \cite{Data:Airplane}.
\\Cloud-top heights data are from the SEVIRI instrument aboard Meteosat Second Generation (MSG) satellite~\cite{Meirink:2022}.

\section*{Acknowledgements}

BMH  ML, and PT are supported by ERC [Grant Agreement No.\ 101041097];
NL acknowledges the AFOSR Award FA9550-24-1-0124 from the University of New Hampshire.

LOFAR~\cite{Haarlem:2013} is designed and constructed by ASTRON collectively operated by the International LOFAR Telescope (ILT) foundation under a joint scientific policy. The ILT resources have benefitted from the following recent major funding sources: CNRS-INSU, Observatoire de Paris and Universit\'{e} d'Orl\'{e}ans, France; BMBF, MIWF-NRW, MPG, Germany; Science Foundation Ireland (SFI), Department of Business, Enterprise and Innovation (DBEI), Ireland; NWO, The Netherlands; The Science and Technology Facilities Council, UK.

\section*{Authors' contributions}

OS performed most of the analysis for this work, wrote the first draft, and made most figures.
ML analyzed the flight data and made the figure of the cloud-top heights and airplane banking angle.
OS, BMH, ML, PT, JD, NL, SC, and CS have participated in the extensive discussions that are at the basis of this work and have contributed to the final manuscript.
StV and BMH were responsible for the data acquisition.
SB, TH, KM, AN, and TNGT contributed a critical reading of the manuscript.

\appendix
\section{The TRI-D beamforming and imaging procedure}\applab{TRID}

When searching for source locations (imaging) using a beamforming technique, the space is voxelated and for each voxel the beam-formed signal is constructed. For a broad-band (impulsive) signal this is done by adding the time traces measured in all antennas while compensating for the travel times of the signal from a particular voxel location to the positions of the individual antennas. The idea is that the time-traces should all interfere constructively (are coherent and thus have maximal intensity for the beam-formed signal) when all time-shifts are correctly accounted for, i.e., the source is at the position of the particular voxel.
A major complication arises when the source is polarized (which it usually is) and the viewing direction of the voxel depends on the antenna position (as it will for near-field imaging). In this case the phase (sign) of the measured signal depends on the viewing direction. Not accounting for this phase makes that the beam-formed signal does not reach full coherency even when pointing to the correct voxel which means that the basic idea behind beamforming fails. It is thus imperative to account for the polarization of the source in the beamforming procedure, which is where the Time Resolved Interferometric 3-Dimensional (TRI-D) imaging procedure was invented for, see Refs.~\cite{Scholten:2021-INL, Scholten:2022}.

In the TRI-D procedure it is assumed that a point-like source emitting polarized dipole radiation is placed at the center of a voxel. The polarization orientation and magnitude of this dipole is considered an unknown parameter and is denoted as a 3D vector $\vec{I}$.
The measured voltages produced by the antenna can be related to the incoming electric radiation $\vec{E}_a$ through the Jones matrix. For sake of simplicity we skip this stage in the present discussion and consider $\vec{E}_a$ the measured electric field at antenna $a$. The radiation field is always oriented perpendicular to the propagation direction from the source to the antenna, $\left( \vec{E}_a \cdot \hat{r}_{as} \right) =0$, where the hat denotes a unit vector. The two perpendicular directions to $\hat{r}_{as}$ are denoted by $\hat{r}_{ak}$ where $k=1,2$. The orientations of these unit vectors are dependent on the antenna position with respect to the source and will thus be different for each of the about 200 antennas used for imaging. Assuming that the measured radiation is emitted by a dipole source, with a current-moment change $\vec{I}$, one can optimize the orientation and magnitude of this source by minimizing the sum of the square difference between the measured and the modeled radiation fields,
\beq
 Q^2= \sum_{a,k} \left[ \left( \vec{E}_a \cdot \hat{r}_{ak} \right) - \frac{\left( \vec{I} \cdot \hat{r}_{ak} \right)}{R_{as}} \right]^2 w_{ak} \;. \eqlab{Qual}
\eeq
In \eqref{Qual} the distance between the source and the antenna is denoted by $R_{as}$ and $w_{ak}$ are weighting constants that depend on the signal-to-noise ratio. Because of the linearity of the system of equations, the solution is given by a relatively simple algebraic matrix equation,
\beq
  \vec{I} = A^{-1} \vec{F}  \;, \eqlab{AIF}
\eeq
which is the central equation used in the TRI-D beamforming procedure. The matrix elements of the $3\times3$ tensor $A$ are given by
\beq
A_{ij}= \sum_{a,k} \left(\hat{r}_{ak,i} \, \hat{r}_{ak,j}\right) \,w_{ak}/R_{as}^2 \;,
\eeq
where $i,j$ label cartesian coordinates and
and
\beq
\vec{F}=\sum_{a,k}  E_{ak}\,\hat{r}_{ak}\,w_{ak}/R_{as} \;, \eqlab{F}
\eeq
is the beam-formed sum of the measured electric fields of all antennas (taking into account the proper time delay due to the travel time from the source to each antenna).

In the TRI-D procedure one thus sums the measured fields in all antennas, where the polarization of the signal is properly taken into account, to reconstruct the magnitude and orientation of a point-like source at the center of each voxel.

\section{Grid Fineness}\applab{F}

When performing TRI-D calculations the choice of the grid spacing, used to voxelate space, is important. On one hand one would like to have a small grid spacing as this increases accuracy, however, this goes at the expense of computer time and computer memory. Essential considerations for selecting a grid is that it is of sufficient fineness such that, when calculating the intensity distribution obtained from beaming (as performed in TRI-D), the maximum is captured by the grid or, stated differently, the grid spacing should be a fraction of the beam spread. If the grid is too coarse, the grid points may fall too far from the maximum to detect it and instead a secondary maximum (at a side beam) may show. Since LOFAR antennas are spread semi-randomly with a roughly logarithmic spacing, the estimate of a reasonable grid spacing becomes a challenge. Additionally, the beam spread may be very different in the three directions (Northward, Eastward, upward). To streamline this we have introduced the `Fineness' parameter that is based on an estimate for the beam spread based of the change of the time-delays over a grid spacing, as used in the beaming calculation.

Assume the grid is centered at a position $\vec{r}_g$ and the antenna coordinates are given by $\vec{r}_a$. The grid spacings are given by three vectors $\vec{\delta}_i$, one for each of the three grid axes. The distance of antenna $a$ to the center of the grid can thus be written as $d_a=|\vec{r}_g-\vec{r}_a|$. Since time differences are essential for beamforming of an impulsive signal, we calculate the difference of the time of arrival in antenna $a$ of a signal emitted from the the center of the grid, and from one grid point removed,
\beq
c \,\Delta t_{a,i}=d_a -|\vec{r}_g+\vec{\delta}_i-\vec{r}_a| \;.
\eeq
Assuming that the signals of all antennas are summed fully coherently for the center of the grid, i.e., added with time delays corresponding to the signal travel time, the signal will not be fully coherent when added in one grid-point removed. The amount of incoherence is given by the spread in arrival times. Since the signal amplitude for a spherically symmetric emitting source falls linearly with distance we thus estimate the change in coherent intensity from the maximum $I_0$ for a neighboring voxel to be proportional to
\beq
F_i \propto \frac{\Delta I_i}{I_0} \equiv \Delta_\nu \frac{ \sqrt{ <\left(\Delta t_{a,i}/d_a\right)^2>_a - \left(<\Delta t_{a,i}/d_a>_a\right)^2}}{<1/d_a>_a} \;,
\eeq
where $<>_a$ implies an averaging over all antennas involved in the observation and $\Delta_\nu$ is the effective frequency bandwidth, taken at $10MHz$, for LOFAR. The grid fineness estimate, $F_i$, should serve only as a rough estimate of the dependence of the Beamforming interference pattern on grid spacing. When the grid spacing is much smaller than the distance to the nearest antenna, i.e., all realistic cases, the grid fineness estimate, $F_i$ is proportional to grid spacing $|\vec{\delta}|_i$ where the proportionality factor depends on the direction $i$. Since in this work we take the same grid fineness constant for all three orientations of the grid, $F_{1,2,3}=F$ this implies a different grid spacing in the three directions.

\section{Polarization}\applab{pol}

In the TRI-D procedure the source polarization is calculated for all time samples in the slicing window. For a subsequent analysis this fine structure is usually too detailed and for this reason a principal component analysis, see \appref{C_PCA}, is applied to the -per sample- polarization orientations within a single pulse. The polarization density matrix is constructed as
\beq
\rho_{i,j}=\sum_t {\cal J}_i(t) {\cal J}_j^*(t) \;, \eqlab{rho-pol}
\eeq
where the sum runs over all time samples in the pulse, ${\cal J}(t)$ is the dipole current moment change for this time sample, and $i,j$ denote cartesian coordinates. The $^*$ denotes complex conjugation. Following Ref.~\cite{Setala:2002}, the 9 real Stokes parameter may be constructed from the Hermitian polarization density matrix, given in \eqref{rho-pol}, as,
\bea
I &=& \rho_{1,1} + \rho_{2,2} + \rho_{3,3} \;, \nonumber \\
I_{12} &=& \rho_{1,1} + \rho_{2,2} \;,  \nonumber \\
Q &=& \rho_{1,1} - \rho_{2,2} \;, \nonumber \\
U &=& \rho_{2,1} + \rho_{1,2} =2 \Re(\rho_{1,2}) \;, \nonumber \\
V &=& i(\rho_{2,1} - \rho_{1,2} ) =2 \Im(\rho_{1,2}) \;, \nonumber \\
U_2 &=& \rho_{1,3} + \rho_{3,1} \;, \nonumber \\
V_2 &=& i(\rho_{1,3} - \rho_{3,1} ) \;, \nonumber \\
U_1 &=& \rho_{3,2} + \rho_{2,3} \;, \nonumber \\
V_1 &=& i(\rho_{3,2} - \rho_{2,3} ) \;,  \\
I_3 &=& \rho_{3,3} =I-I_{12} \;, \nonumber \\
I_8 &=& (I_{12}-2I_3)/\sqrt{3} \;, \nonumber \eqlab{Stokes}
\eea
where the last two are not independent and $i$ is the imaginary unit. In should be noted that the diagonal matrix elements of an Hermitian matrix are real-valued.

The Stokes parameters contain the complete information on the total polarization of the source, however, this information is not very transparent. Following Ref.~\cite{Setala:2002} it is instructive to extract the un-, linearly-, and circularly-polarized fractions of the source as
\bea
   P_{\rm un} &=& 1 - P_{\rm lin} - P_{\rm circ} \;,\nonumber \\
   P_{\rm lin} &=& \frac{3}{4\,I^2}\left( Q^2 + U^2 + U_1^2 + U_2^2 +I_8^2\right) \;,\\ \eqlab{P_lin}
   P_{\rm circ} &=& \frac{3}{4\,I^2}\left( V^2 + V_1^2 + V_2^2 \right)  \;. \nonumber
\eea
For a simple dipole source the fraction of linear polarization should be unity. However, a polarization direction that is gradually changing over the duration of the source, gives rise to a circularly polarized component~\cite{Scholten:2016}. A polarization direction that varies randomly is reflected in the unpolarized fraction.

\subsection{Complex principal component polarization analysis}\applab{C_PCA}

An alternative, more intuitive, way to express the polarization of the source is by performing a principal component analysis on the polarization density matrix. The orientation of the three axes are obtained from the complex eigenvectors $\vec{\epsilon}_j$ of the polarization density matrix \eqref{rho-pol} together with the real-valued eigenvalues $\lambda_j$. Care should be taken with the interpretation of the eigenvectors as they may be multiplied by an arbitrary phase $e^{i\phi}$. We use this to calculate the phase $\phi_j$ that maximizes the length of the real part of each eigenvector $\vec{\epsilon}_j$,
\beq
e^{i2\phi_j}=\frac{\vec{\epsilon}_j \cdot \vec{\epsilon}_j}{|\vec{\epsilon}_j \cdot \vec{\epsilon}_j |} \;, \eqlab{phase}
\eeq
where ${\vec{\epsilon}_j \cdot \vec{\epsilon^*}_j}=1$ for a properly normalized eigenvector. Multiplying the eigenvector with the phase factor we obtain the form with a maximal real component,
\beq
\vec{\epsilon^\prime}_j=e^{-i\phi_j} \vec{\epsilon}_j  \;. \eqlab{prime}
\eeq
For our case, the physical interpretation of $\vec{\epsilon^\prime}_j$ (normalized to unity) is that the
current-moment change, $\vec{I}$, extracted from the data using \eqref{AIF},
is oscillating (with magnitude $\sqrt{\lambda_j}$) out of phase in the $\Re(\vec{\epsilon^\prime}_j)$ and $\Im(\vec{\epsilon^\prime}_j)$ directions. Maximizing the real part is equivalent to orthogonalizing $\Re(\vec{\epsilon^\prime}_j)$ and $\Im(\vec{\epsilon^\prime}_j)$. $\vec{\epsilon^\prime}_j$ can thus be viewed as a superposition of circular and linear polarizations as discussed in the following section.

\subsection{Analyzing the eigenvectors}

Each eigenvector is complex and we order the eigenvalues such that $\lambda_1>\lambda_2>\lambda_3$. To analyze each eigenvector we assume (as discussed in the previous section) that the $j^{th}$ eigenvector can be written, dropping the index $_j$, as
\beq
\vec{\epsilon^\prime}= \vec{A} +i\vec{B} \;,
\eeq
with $\vec{A} \cdot \vec{B}=0$, achieved by maximizing the real component. The normalization of $\vec{\epsilon^\prime}$ implies $|A|^2+|B|^2=1$. In the basis spanned by $\hat{A}$, $\hat{B}$, and $\hat{C}=\hat{A}\times \hat{B}$ where $|A|\hat{A}=\vec{A}$, etc.,
we can construct a pure normalized circularly polarized column vector as
\beq
\hat{\epsilon}_{\rm circ}= \hat{A}/\sqrt{2} + i\hat{B}/\sqrt{2} \;,
\eeq
yielding the polarization density
\beq
\rho^c = \hat{\epsilon}_{\rm circ} \hat{\epsilon}_{\rm circ}^\dagger = \frac{1}{2}\left[ {\begin{array}{ccc}
   1 & -i & 0 \\
   i & 1 & 0 \\
   0 & 0 & 0 \\
  \end{array} } \right] \;.\eqlab{rho-circ}
\eeq
Likewise we may construct a pure linearly polarized column vector as
\beq
\hat{\epsilon}_{\rm lin}= \hat{A} \;,
\eeq
yielding the polarization density
\beq
\rho^l = \hat{\epsilon}_{\rm lin} \hat{\epsilon}_{\rm lin}^\dagger = \left[ {\begin{array}{ccc}
   1 & 0 & 0 \\
   0 & 0 & 0 \\
   0 & 0 & 0 \\
  \end{array} } \right] \;.\eqlab{rho-lin}
\eeq
We can now rewrite
\beq
\vec{\epsilon^\prime}= |A|\hat{A} +i|B|\hat{B}
   =\left(|A|-|B|\right) \hat{\epsilon}_{\rm lin} + \sqrt{2}|B|\,\hat{\epsilon}_{\rm circ} \;,
\eeq
yielding the polarization density
\beq
\rho = \vec{\epsilon^\prime} \vec{\epsilon^\prime}^\dagger = \left[ {\begin{array}{ccc}
   |A|^2 & -i|A||B| & 0 \\
   i|A||B| & |B|^2 & 0 \\
   0 & 0 & 0 \\
  \end{array} } \right] \;. \eqlab{rho-ev}
\eeq
Using the standard expressions for the fractions of circular and linear polarization in 2D while using
$I^2=(|A|^2+|B|^2)^2=1$, we obtain $P_{\rm circ}=V^2/I^2=4|A|^2|B|^2$, and , $P_{\rm lin}=Q^2/I^2=(|A|^2-|B|^2)^2$, where $P_{\rm circ}+P_{\rm lin}=(|A|^2+|B|^2)^2=1$. The linear polarization is oriented along $\hat{A}$ while the direction of circular polarization is $\hat{C}$. Note that the polarization density for this eigenvector, \eqref{rho-ev}, cannot be written as the sum (with suitably chosen values) of the circular, \eqref{rho-circ}, and linear densities, \eqref{rho-lin}.

Alternative definitions for the linear and circular polarization fractions may be obtained by introducing the angle between $\Re(\vec{\epsilon^\prime})$ and $\vec{\epsilon^\prime}$ as
\beq
\omega  = \arccos{|\Re(\vec{\epsilon^\prime})|} = \arccos{|A|} \;. \eqlab{domg}
\eeq
The linear polarization probability can now be written as
\beq
 P_{\rm lin}=(|A|^2-|B|^2)^2 = (\cos^2{\omega} -\sin^2{\omega})^2=\cos^2{2 \omega} \;,
\eeq
and
\beq
P_{\rm circ} = \sin^2{2 \omega} \;. \eqlab{P_circ_PCA}
\eeq
The length of the linear and circular polarization vectors are thus $\lambda \cos{2 \omega}$ and $\lambda \sin{2 \omega} $ respectively.

\subsection{Discussion}

Through principal component analysis we have separated the 3D-polarization problem into 3 incoherent 2D problems. Incoherent in this sense means that the original pulse, giving rise to the complete polarization density is separated into 3 separate pulses that are assumed to have no overlap in time and thus do not interfere. This is expressed by writing the polarization matrix as a direct sum of three separate polarization densities, each corresponding to an eigenvector $\vec{\epsilon}_j$ with eigenvalue $\lambda_j$. For each component the directions and magnitudes of circular and linear polarization can be determined as discussed in the previous section.

To understand the relation between the principal component analysis and the unpolarized fraction introduced in \eqref{P_lin} it is simplest to consider the case where the eigenvectors are real. In that case, taking a basis that is lined-up with the eigenvectors, the Stokes parameters,  \eqref{Stokes}, relate directly to the eigenvalues $\lambda_i$ as $I=\sum_i \lambda_i$, $I_{12}=\lambda_1+ \lambda_2$, $Q=\lambda_1- \lambda_2$, $I_3=\lambda_3$, and $I_8=(\lambda_1+ \lambda_2-2\lambda_3)/\sqrt{3}$ with all others zero. This yields
\bea
P^r_{\rm un} &=& \frac{3}{I^2}\left( \lambda_1\lambda_2 + \lambda_1\lambda_3 + \lambda_2\lambda_3\right)  \;,\nonumber \\ \eqlab{P_linr}
   P^r_{\rm lin}
&=& 1-P^r_{\rm un} \\
   P^r_{\rm circ} &=& 0 \;. \nonumber \eqlab{PrLin}
\eea
The extreme case $\lambda_2 = \lambda_3=0$ gives $P^r_{\rm lin}=1$ as expected and $\lambda_1 = \lambda_2$ with $\lambda_3=0$ gives $P^r_{\rm lin}=1/4$. One thus should not add the linear or circularly polarized parts in each direction to get the polarized fractions.

\bibliography{}

\end{document}